\documentclass[twocolumn]{autart}    
\usepackage{algorithm}
\usepackage{algorithmic}

\usepackage{amsmath} 
\usepackage{esint}
\usepackage{graphicx}          
\usepackage{color}                     
\usepackage{amssymb}
\usepackage{cuted}
\usepackage{lipsum} 
\allowdisplaybreaks[1]
\newtheorem{proposition}{Proposition}
\newtheorem{remark}{Remark}

\usepackage[outdir=./]{epstopdf}
\usepackage{longtable}
\usepackage{multirow} 
\usepackage{multicol}

\usepackage{bm}
\usepackage{float}
\usepackage{color}
\usepackage{amssymb}
\usepackage{booktabs}
\usepackage{mathrsfs}
\usepackage{color}
\usepackage{caption}
\usepackage{subfig}
\usepackage{enumerate}
\usepackage{amsmath} 
\usepackage{cite} 
\usepackage{setspace}

\usepackage{natbib}

\begin{document}

\begin{frontmatter}

\title{A Variational Bayes Moving Horizon Estimation \\Adaptive  Filter\ with Guaranteed Stability \thanksref{footnoteinfo}} 

\thanks[footnoteinfo]{This paper was not presented at any IFAC 
meeting. Corresponding author: Yunze Cai. 
}

\author[Shanghai1,Shanghai2]{Xiangxiang Dong}\ead{js.danesir@sjtu.edu.cn},    
\author[Florence]{Giorgio Battistelli}\ead{giorgio.battistelli@unifi.it},              
\author[Florence]{Luigi Chisci}\ead{luigi.chisci@unifi.it},  
\author[Shanghai1,Shanghai2]{Yunze Cai}\ead{yzcai@sjtu.edu.cn}

\address[Shanghai1]{Department of Automation, Shanghai Jiao Tong University, Shanghai, 200240, China}
\address[Shanghai2]{Key Laboratory of System Control and Information Processing, Ministry of Education of China, Shanghai 200240, China}
\address[Florence]{Department of Information Engineering, University of Florence, Florence, 50139, Italy}


\begin{abstract}                          
This paper addresses state estimation of linear systems 
with special attention on unknown process and measurement noise covariances, 
aiming to enhance estimation accuracy while preserving the stability guarantee of the Kalman filter. 
To this end, 
the full information estimation problem over a finite interval is firstly  addressed. Then,
a novel adaptive
\textit{variational Bayesian} (VB) \textit{moving horizon estimation} (MHE) method is proposed, 
exploiting VB inference, MHE and Monte Carlo integration with importance sampling
for joint estimation of the unknown process and measurement noise covariances,
along with the state trajectory over a moving window of fixed length.
Further, it is proved that the proposed adaptive VB MHE filter ensures mean-square boundedness 
of the estimation error 
with any number of importance samples and VB iterations, as well as for any window length.
Finally, simulation results on a target tracking example 
demonstrate the effectiveness of the VB MHE filter 
with enhanced estimation accuracy and convergence properties
compared to the conventional non-adaptive Kalman filter and other existing adaptive filters.
\end{abstract}

\end{frontmatter}

\section{Introduction}
\label{intro}

The Kalman filter (KF) is an optimal filter for state estimation of linear stochastic dynamical systems \citep{10.1115/1.3662552}.
It has been widely exploited in a variety of applications (e.g., signal processing, target tracking, control systems, etc.) 
in view of its ease of implementation and strong exponential stability properties \citep{10.5555/863669,6371743,845941}. 

Even though the conventional KF performs state estimation with guaranteed stability, 
its performance is highly affected by 
the prior knowledge on process and measurement noises,  
which are typically assumed to have Gaussian distributions with known \textit{process noise covariance matrix} (PNCM) 
and \textit{measurement noise covariance matrix} (MNCM) \citep{7929421,1100100}.
Unfortunately, in practical situations the noise statistics are usually unknown or only partially known, due to partial and/or imprecise prior knowledge.
To address this issue, many adaptive methods have been proposed  \citep{1100100}, 
including covariance matching, maximum likelihood  and \textit{variational Bayesian} (VB),   
where the VB approach \citep{7929421,4644060,8588399} is commonly used for joint estimation of state and unknown noise statistics
in view of its high estimation accuracy.

An adaptive VB filter is presented by \citet{8025799} 
where the \textit{predicted error covariance matrix} (PECM) and MNCM are jointly estimated together with the system state.
While effective in many contexts, its filtering performance can be sensitive to the choice of the nominal process noise covariance, 
which is used to calculate the initial value of the PECM. 
The approach of \citet{8025799}  has been extended and applied also in nonlinear, multi-sensor, 
and multimodal settings
 \citep{DONG2021107837,SPL,9042265}.
As an improvement, a \textit{variational Bayes sliding window Kalman filter} (VB Sliding Window) 
is proposed in \citep{9096408} by imposing an approximation on the smoothing posterior 
\textit{probability density function} (PDF)
of the sliding window states. 
However, the performance of this filter is affected by the window length
and, more importantly, 
for both adaptive filters in \citep{8025799} and \citep{9096408} no theoretical guarantee of stability of the estimation error has been proved.

Besides  the KF and its variants, another widely employed estimation technique is
moving horizon estimation (MHE) \citep{RAO20011619,1184901}.
MHE is based on the idea of computing an estimate of the state trajectory over a moving window of fixed length
by taking into account a limited amount of most recent information, 
after which  the estimation results 
are propagated to the next time step and then the former stated estimation procedure is further repeated.
The main positive feature of MHE is the possibility of defining a performance criterion that can be designed specifically for the problem under consideration.
Thanks to its guaranteed stability and performance \citep{1178905,ALESSANDRI20081753}, MHE has been widely used in both
linear and nonlinear contexts \citep{BATTISTELLI2017374,7835659,9302761,8967155,ALESSANDRI201685} for centralized, networked, and distributed estimation
\citep{YIN2017152,LAURICELLA,ZOU2020109154,5437313,8523657,FARINA2010910,PDMHE,6473849}.
The interested reader is referred to the special issue \citep{ALBA} for recent advances on MHE.
In a Bayesian framework, 
MHE can be conveniently exploited to approximate the full-information Bayesian estimation problem 
whenever the latter does not admit a closed-form recursive solution
 \citep{DELGADO20141116,9143645}.

In this work,  in order to perform joint estimation of state, PNCM and, MNCM with bounded estimation error, 
a novel adaptive VB MHE filter 
is developed. 
To this end, inspired by the idea of MHE, the unknown noise covariances 
are regarded as nearly constant within the current window
and estimated through the VB method while ensuring available bounds. The considered 
framework allows for imposing constraints on the unknown PNCM and MNCM in terms of 
a priori defined sets
to which the respective estimates should belong.
 First, the full information estimation problem over a finite interval is addressed by modeling the PNCM and MNCM distributions
as constrained inverse Wishart. In this context, we provide an algorithm based on the VB fixed-point method for computing the optimal 
factorized approximation of the joint posterior of state trajectory, PNCM and MNCM.
Then, the MHE paradigm is employed to make the proposed approach recursive and ensure bounded memory and computational complexity.
The resulting VB MHE filter exploits Monte Carlo integration with importance sampling 
for computing the estimates of the unknown covariances. 
Further, and most importantly, it is proved that the proposed VB MHE filter ensures stability, in terms of mean-square boundedness of the estimation error, 
for any choice of the number of importance samples and VB iterations, as well as of the window length.
Simulation results demonstrate the effectiveness of the proposed filter as compared to the state of the art, thus confirming the theoretical findings.

The remaining parts of this paper are organized as follows.  
Section \ref{probm} provides background and the problem formulation. 
{In Section \ref{non-bounded-MHE}, the full information estimation problem over a finite interval is addressed
via VB inference. Then, in Section \ref{bounded-MHE} the MHE paradigm is applied to derive a recursive adaptive
estimation algorithm.}
In Section \ref{stability}, the stability of the proposed adaptive filter is analyzed.
Performance assessment via simulation experiments concerning a target tracking example is provided
in  Section \ref{simulation}.
Finally, conclusions and perspectives for future work are given in Section \ref{conclusion}.

\section{Problem formulation and preliminaries}
\label{probm}

Consider a linear discrete-time system 
\begin{eqnarray}
x_t &=& A x_{t-1} + w_{t-1}  
\label{state-eq}
\end{eqnarray}
and linear measurements 
\begin{eqnarray}
y_t &=& C x_t + v_t
\label{measurement-eq}
\end{eqnarray}
where: 
$t$ is the time index;
$x_t$ and $y_t$ are the state and measurement vectors of dimensions $n_x$ and $n_y$, respectively; 
$A$ and $C$ are the state transition and, respectively, measurement matrices; 
$w_{t-1}$ and $v_t$ denote white process and measurement noises, assumed to be Gaussian 
with zero mean  but unknown 
covariances $Q$ and $R$.
It is also assumed that $w_k$ and $v_j$ are uncorrelated for any $k$ and $j$.
The unknown covariances  $Q$ and $R$ are supposed to belong to known sets $\mathcal Q \subseteq \mathbb{S}^{n_x}_+$ and $\mathcal R \subseteq \mathbb{S}^{n_y}_+$, respectively,
where $\mathbb{S}^d_+$ denotes the set of real-valued positive definite $d \times d$  symmetric matrices.

{Following a Bayesian approach,  
the process and measurement noise covariances are regarded as random matrices to be estimated
together with the state trajectory. For the resulting adaptive estimation problem,} 
although there have been some variational adaptive filters proposed in \citep{8025799} and \citep{9096408},
for such filters there is no available proof of stability. Motivated by this,
this paper aims to propose a novel adaptive filter for unknown PNCM and MNCM 
that ensures mean-square boundedness of the estimation error. 
The main contribution focuses on the derivation of the proposed filter as well as on the stability proof. 
The proposed adaptive filter will jointly exploit VB inference and MHE.


\subsection{Idea of variational Bayes inference}

Before deriving the proposed adaptive filter with guaranteed stability, 
this section briefly recalls the basic idea of VB inference. 

{The VB approach is based on the idea of approximating the true posterior $p(\theta)$
with a variational distribution $q(\theta)$ constrained to have a fixed form
by minimizing the  \textit{Kullback-Leibler divergence} (KLD) 
from  $p(\theta)$ \citep{4644060}, i.e., 
\begin{eqnarray}
q = \arg \min_{q}\, \textrm{KLD} \left(q||p\right)
\label{VB-idea}
\end{eqnarray}
where KLD is defined as 
\begin{eqnarray}
\textrm{KLD} \left(q||p\right) = \int q\left(\theta\right) 
\log \frac{q\left(\theta\right)}{p\left(\theta\right)}\, d \theta  \,.
\label{eq:KLD}
\end{eqnarray}
When the variables to be estimated can be partitioned as $\theta = (\theta_1, \ldots, \theta_M)$ and the variational distribution $q(\theta)$ is given the factorized form $q(\theta) =  \prod_{i=1}^{M} q_i (\theta_i)$, then the optimal solution to
(\ref{VB-idea}) must satisfy
\begin{equation}\label{eq:VB}
\log q_i (\theta_i) = \mathop{E}_{ \theta_j , j \ne i } [ \log p(\theta) ] + {\rm constant}
\end{equation}
where $E$ denotes expectation.
Both  VB and VB sliding-window filters of  \citep{8025799} and \citep{9096408} 
have been derived via VB inference in factorized form.}

\section{ Variational Bayes inference for full information  estimation}
\label{non-bounded-MHE}

{In this section, the full information estimation problem over a finite interval is addressed and
VB inference is used in order to compute a factorized approximation of the true joint posterior of the state trajectory, PNCM, and MNCM.}

For the system (\ref{state-eq})-(\ref{measurement-eq}),
the initial state $x_0$ is assumed to be Gaussian-distributed 
with mean $\bar{x}_0$ and covariance $\bar{P}_0$, i.e., 
\begin{eqnarray}
p(x_0) &=& \mathcal{N} \left( x_0; \bar{x}_{0}, \bar{P}_{0} \right) .
\label{init-x}  
\end{eqnarray}
For the unknown covariance of a Gaussian distribution, 
its conjugate prior is the inverse Wishart distribution \citep{8025799,soton46376}, 
whose PDF is denoted as $\mathcal{W}^{-1} \left(G;\Delta, \gamma \right)$, indicating that the
random matrix { $G \in  \mathbb{S}^d_+$} follows an inverse Wishart distribution 
with degree of freedom $\gamma > d + 1$ and scale matrix { $\Delta \in \mathbb{S}^d_+$}.
{Since the unknown covariances $Q$ and $R$ are supposed to belong to the sets $\mathcal Q$ and $\mathcal R$, respectively,
we take the priors for $Q$ and $R$ as constrained  inverse Wishart distributions of parameters $(\bar{M},\bar{m})$ and, respectively, $(\bar{S}$, $\bar{s})
$, i.e.
\begin{eqnarray}
p\left(Q\right) &\propto& \mathcal{W}^{-1} \left( Q; \bar{M}, \bar{m} \right) \,  \mathbf{1}_{\mathcal{Q}}\left(Q\right)  \label{prior-Q-bounded}    \\
p\left(R\right) &\propto& \mathcal{W}^{-1} \left( R; \bar{S}, \bar{s} \right) \, \mathbf{1}_{\mathcal{R}}\left(R\right) 
\label{prior-R-bounded}
\end{eqnarray}
where $\mathbf{1}_{\mathcal{Q}} (Q)$ is  the indicator function taking value $1$ if $Q \in \mathcal{Q}$ and $0$ otherwise.
}


Let $y_{1:t}$ denote the sequence of measurements from time $1$ to time $t$.
Then, it is an easy matter to check that
the joint posterior PDF of the state trajectory $x_{0:t}$ and the unknown noise covariances $Q, R$ can be expressed as 
\begin{eqnarray}
&& p\left( x_{0:t}, Q, R \,|\, y_{1:t}\right) \nonumber \\
&&\propto \mathcal{N} \left( x_0; \bar{x}_{0}, \bar{P}_{0} \right) 
\times \prod_{i=1}^{t} \mathcal{N} \left( x_i; A{x}_{i-1}, Q \right)  \mathcal{N} \left( y_i; C{x}_{i}, R \right)
\nonumber \\
&&\times \mathcal{W}^{-1} \left( Q; \bar{M}, \bar{m} \right)  \, \mathcal{W}^{-1} \left( R; \bar{S}, \bar{s} \right) \, 
 { \mathbf{1}_{\mathcal{Q}}\left(Q\right)  \,  \mathbf{1}_{\mathcal{R}}\left(R\right) }  \, .
\label{joint-PDF}
\end{eqnarray}
By exploiting the VB approach, a factorized approximation of the joint PDF in (\ref{joint-PDF}) is sought as
\begin{eqnarray}
p\left( x_{0:t}, Q, R \,|\, y_{1:t}\right) \cong q_x\left(x_{0:t}\right) q_Q\left(Q\right) q_R\left(R\right)
\label{factored-PDF}
\end{eqnarray}
where $q_x , \, q_Q , \, q_R$ denote the factors of the approximated joint PDF.
To this end, we notice preliminarily that,
since the joint posterior (\ref{joint-PDF}) is null when $Q \notin \mathcal{Q}$ or $R \notin \mathcal{R}$,
then in the VB approximation 
it must hold that
\begin{eqnarray} 
\begin{cases} 
q_Q\left(Q\right) = 0 &\mathrm{for} \, Q \notin \mathcal{Q} \\
q_R\left(R\right) = 0 &\mathrm{for} \, R \notin \mathcal{R}  \, .
\label{case0-bounded}
\end{cases}
\end{eqnarray}
{Otherwise the minimum in (\ref{VB-idea}) could not be achieved since, by definition,
the KLD is infinite whenever the support of $q$ is not contained in the support of $p$.

Concerning the variational distribution of the state trajectory, the following result holds.
\begin{proposition}
\label{proposition-1}
Given the joint posterior (\ref{joint-PDF}) and the factorized approximation (\ref{factored-PDF}), then
the approximated PDF of the state trajectory according to the VB approach is of the form
\begin{eqnarray}
&&q_x \left( x_{0:t} \right)   =  \mathcal{N} \left(  x_{0:t}; \hat x \left({\Psi},{\Phi} \right), 
P \left({\Psi},{\Phi}\right)  \right)
\label{distri-x}
\end{eqnarray}
where
\begin{eqnarray}
&&\Phi = \int_{\mathcal Q} Q^{-1} \, q_Q\left(Q\right) \, dQ    \label{est-Q} \\
&&\Psi = \int_{\mathcal R} R^{-1} \, q_R \left(R\right) \, dR    \label{est-R} \\
&&  \hat{x}  \left(\Psi,\Phi\right)  =  \Omega^{-1}\left(\Psi,\Phi \right) \omega \left(\Psi \right) \label{x-mean} \\
&& P \left(\Psi, \Phi \right) = \Omega^{-1} \left(\Psi, \Phi \right) \label{x-P} \\
\label{P}
&& \omega \left(\Psi \right) = 
\begin{bmatrix} C' \Psi y_t   \\    \vdots    \\   C' \Psi y_1   \\   
\bar{P}_0^{-1}\bar{x}_0 
\end{bmatrix}
\label{omega}
\end{eqnarray} 
and the block matrix \, $\Omega \left(\Psi, \Phi \right)$ defined as in equation (\ref{Omega}).
\begin{figure*}[tb]
\hrulefill
\begin{equation}
\Omega \left(\Psi,\Phi \right) =  \begin{bmatrix}  
&C'\Psi C+ \Phi  &-\Phi A  &0   &\cdots  &0   \\
&-A'\Phi  &C'\Psi C +\Phi + A'\Phi A   &-\Phi A  & &\vdots   \\
&0 &-A' \Phi  &\ddots  &\ddots  &\vdots\\
&\vdots  &   &  &C'\Psi C+\Phi + A'\Phi A  &-\Phi A  \\
&0   &\cdots  &\cdots   &-A'\Phi  &A'\Phi A+\bar{P}^{-1}_{0}     \end{bmatrix} 
\label{Omega}
\end{equation}
\hrulefill
\end{figure*}
\end{proposition}


{\bf Proof.\,} 
}
By taking logarithm on both sides of (\ref{factored-PDF}), we obtain
\begin{eqnarray}
&\mathrm{log}& \, p\left( x_{0:t}, Q, R \,|\, y_{1:t}\right) \nonumber \\
&=& -\frac{1}{2} \bigg[ \Vert x_0 - \bar{x}_{0} \Vert_{\bar{P_0}^{-1}}^2  
+ \sum_{i=1}^{t} \Big( \mathrm{log} |R| +  \Vert y_i - C{x}_{i} \Vert_{R^{-1}}^2 \nonumber \\
&&  + \mathrm{log} |Q| +  \Vert x_i - A{x}_{i-1} \Vert_{Q^{-1}}^2  \Big) + \left( \bar{s}+n_y+1 \right) \mathrm{log} |R| \nonumber \\
&& +  \left( \bar{m}+n_x+1 \right)\mathrm{log} |Q| + \mathrm{tr}\left({\bar{S}R^{-1}}\right)     + \mathrm{tr}\left({\bar{M}Q^{-1}}\right)  \bigg]  \nonumber \\
&& + \, \mbox{constant}  \nonumber \\
&=& -\frac{1}{2} \bigg\{  \left(x_0-\bar{x}_{0}\right)' \bar{P_0}^{-1} \left(x_0-\bar{x}_{0}\right) + \sum_{i=1}^{t} \Big[ (y_i-C{x}_{i})' \nonumber \\
&& \times {R}^{-1} (y_i-C{x}_{i})  + (x_i-A{x}_{i-1})' {Q}^{-1}  (x_i-A{x}_{i-1})  \Big]   \nonumber \\ 
&& + \left( \bar{s}+t+n_y+1 \right) \mathrm{log}|R| + \left( \bar{m}+t+n_x+1 \right) \mathrm{log}|Q|   \nonumber \\
&& + \mathrm{tr}\left( \bar{S}R^{-1}\right)  + \mathrm{tr}\left( \bar{M}Q^{-1}\right) \bigg\}  + \mbox{constant}
\label{log-joint-PDF}
\end{eqnarray}
{for any $Q \in \mathcal Q$ and $R \in \mathcal R$.
Hence, in view of (\ref{eq:VB}) and  (\ref{case0-bounded}), it can be obtained that 
\begin{eqnarray}
&q_x& \left(x_{0:t}\right) \nonumber  \\ 
&\propto& \exp \int_{\mathcal R} \int_{\mathcal Q} \mathrm{log} \, p\left( x_{0:t}, Q, R | y_{1:t}\right) 
q_Q\left(Q\right) q_R\left(R\right)  dQ  dR  \, .  \nonumber \\
\label{Approxi-x}
\end{eqnarray}
In turn, we have
\begin{eqnarray}
&&\int_{\mathcal R} \int_{\mathcal Q}  \mathrm{log} \, p\left( x_{0:t}, Q, R \,|\, y_{1:t}\right) 
q_Q\left(Q\right) q_R\left(R\right) \, dQ \, dR    \nonumber \\
&&=-\frac{1}{2} \bigg\{  \left(x_0-\bar{x}_{0}\right)' \bar{P_0}^{-1} \left(x_0-\bar{x}_{0}\right)  \nonumber \\
&& \quad + \sum_{i=1}^{t} \Big[ (y_i-C{x}_{i})'\Psi (y_i-C {x}_{i})   \nonumber \\
&& \quad + (x_i-A{x}_{i-1})' \Phi (x_i-A{x}_{i-1})  \Big] \bigg\}   + \, {\rm constant}  
\label{integ1-x}
\end{eqnarray}
with $\Phi$ and $\Psi$ defined as in  (\ref{est-Q})-(\ref{est-R}).
Then, with standard algebraic manipulations, it can be checked that the 
approximated PDF of the state trajectory has Gaussian distribution with mean and covariance
given by (\ref{x-mean}) and (\ref{x-P}), respectively.
\hfill $\square$
}

We remark that, by construction, the estimate and covariance of Proposition 1 are partitioned as follows 
\begin{eqnarray}
\hat{x} =  \begin{bmatrix}  \hat{x}_t \\  \hat{x}_{t-1}\\   \vdots \\  \hat{x}_0   \end{bmatrix},
\; 
P =  \begin{bmatrix}  &P_t  &P_{t,t-1}  &\cdots  &P_{t,0}   \\
&P_{t-1,t}  &P_{t-1}  &\cdots &P_{t-1,0}  \\
&\vdots &\vdots  &\ddots  &\vdots  \\
&P_{0,t} &P_{0,t-1} &\cdots  &P_{0}
\end{bmatrix}  .
\label{matrix-P}
\end{eqnarray}

{ Concerning the variational distributions of the PNCM and MNCM, the following result holds.}

\begin{proposition}
\label{proposition-2}
Given the joint posterior (\ref{joint-PDF}) and the factorized approximation (\ref{factored-PDF}), then
the approximated PDFs of the PNCM and MNCM according to the VB approach are of the form
\begin{eqnarray}
q_Q \left(Q\right) &\propto& \mathcal{W}^{-1} \left( Q; M\left(\hat{x},P \right), m \right)   \, \mathbf{1}_{\mathcal{Q}} (Q)
\label{distri-Q} \\
q_R \left(R\right) &\propto& \mathcal{W}^{-1} \left( R; S\left(\hat{x},P\right), s \right) \,  \mathbf{1}_{\mathcal{R}} (R)
\label{distri-R}
\end{eqnarray}
with
\begin{eqnarray}
&&m = \bar{m} + t  \label{Q-m} \\
&&s = \bar{s} + t   \label{R-s}  \\
&&M\left( \hat{x},P\right) = \bar{M} + \sum_{i=1}^{t} \Big[ \left(\hat{x}_i-A\hat{x}_{i-1}\right) \left(\hat{x}_i-A\hat{x}_{i-1}\right)'    \nonumber \\
&& \quad\quad\quad\qquad +  P_{i}  + AP_{i-1}A' - P_{i,i-1}A'  - AP_{i-1,i}  \Big]  
\label{Q-M} \\
&&S\left(\hat{x},P\right) = \bar{S} + \sum_{i=1}^{t} \Big[ \left(y_i-C\hat{x}_{i}\right) 
\left(y_i- C\hat{x}_{i}\right)'  \nonumber \\ 
&& \quad\qquad\qquad + C P_{i} C'   \Big] \, .
\label{R-S}
\end{eqnarray}
\end{proposition}
{\bf Proof.\, }  As already discussed, $q_Q\left(Q\right) = 0$ and $q_R\left(R\right) = 0$ for $Q \notin \mathcal Q$ and $R \notin \mathcal R$, respectively.
For $Q \in \mathcal Q$, 
we have 
\begin{eqnarray}
&&q_Q\left(Q\right) \propto  \nonumber \\
&& \exp \int_{\mathcal R} \int \mathrm{log} \, p\left( x_{0:t}, Q, R | y_{1:t}\right)  
q_x(x_{0:t}) q_R(R) \, dx_{0:t}  dR  \nonumber \\  
\label{Approxi-Q}
\end{eqnarray}
where 
\begin{eqnarray}
&&\int_{\mathcal R} \int\mathrm{log} \, p\left( x_{0:t}, Q, R \,|\, y_{1:t}\right)  
q_x(x_{0:t}) q_R(R) \, dx_{0:t}  dR    \nonumber \\ 
&&= -\frac{1}{2} \bigg\{  \left(\bar{m}+n_x+t+1\right) \mathrm{log} \,Q + \mathrm{tr}\left(\bar{M}Q^{-1}\right) \nonumber \\ 
&& \quad + \sum_{i=1}^{t} \mathrm{tr}\Big[ \int \left(x_i-A{x}_{i-1}\right) \left(x_i-A{x}_{i-1}\right)' 
\nonumber \\  
&& \quad \times q_x(x_{0:t})\,  dx_{0:t} \,\,Q^{-1} \Big] \bigg\}  + \, {\rm constant}  \, .
\label{integ1-Q}
\end{eqnarray}
Each integral in the summation turns out to be
\begin{eqnarray}
&& \int \left(x_i-A{x}_{i-1}\right) \left(x_i-A{x}_{i-1}\right)'  q_x(x_{0:t}) \, dx_{0:t}  \nonumber \\ 
&&=  \left(\hat{x}_i-A\hat{x}_{i-1}\right) \left(\hat{x}_i-A\hat{x}_{i-1}\right)'+ AP_{i-1}A'   \nonumber \\ 
&&\quad  +  P_{i}  - P_{i,i-1}A'  - AP_{i-1,i}
\label{integ2-Q}
\end{eqnarray}
where $P_{i}$ are the diagonal blocks of $P$  in (\ref{matrix-P}), 
and the cross-covariances $P_{i-1,i}$ are the lower diagonal blocks 
of $P$.
Then the approximated PDF of the PNCM follows  a bounded inverse Wishart distribution
as in (\ref{distri-Q}).
Further, for $R \in \mathcal R$, we have
\begin{eqnarray}
&&q_R(R) \propto  \nonumber \\
&& \exp \int_{\mathcal Q} \int \mathrm{log} \, p\left( x_{0:t}, Q, R \,|\, y_{1:t}\right)  
q_x(x_{0:t}) q_Q(Q)  dx_{0:t}  dQ     \nonumber \\
\label{Approxi-R}
\end{eqnarray}
where 
\begin{eqnarray}
&&\int_{\mathcal Q} \int  \mathrm{log} \, p\left( x_{0:t}, Q, R \,|\, y_{1:t}\right)  
q_x(x_{0:t}) q_Q(Q)  dx_{0:t}  dQ     \nonumber \\ 
&&=  -\frac{1}{2} \bigg\{  \left(\bar{s}+n_y+t+1\right) \mathrm{log}R + \mathrm{tr}\left(\bar{S}R^{-1}\right) \nonumber \\ 
&& \quad + \sum_{i=1}^{t} \mathrm{tr}\Big[ \int \left(y_i-C{x}_{i}\right) \left(y_i-C{x}_{i}\right)' 
\nonumber \\  
&& \quad \times q_x(x_{0:t})\,  dx_{0:t} \,\,R^{-1} \Big] \bigg\}  + \, {\rm constant} \, .  
\label{integ1-R}
\end{eqnarray}
Each integral in the summation turns out to be
\begin{eqnarray}
&&\int  \left(y_i-C{x}_{i}\right) \left(y_i-C{x}_{i}\right)'   q_x(x_{0:t}) \, dx_{0:t}  \nonumber \\ 
&&=  \left(y_i-C \hat{x}_{i}\right) \left(y_i-C\hat{x}_{i}\right)' + C P_{i}C'  \, .
\label{integ2-R}
\end{eqnarray}
Then the approximated PDF of the MNCM follows 
a bounded inverse Wishart distribution of the form (\ref{distri-R}). \\
%
\mbox{} \hfill $\square$

{
By combining Propositions \ref{proposition-1} and \ref{proposition-2}, 
we get a system of nonlinear equations in the unknowns 
$\hat{x}$, $P$, $M$ and $S$ which can be iteratively solved via the fixed-point method
 estimating one parameter at a time while fixing the others \citep{8025799}. See \citep{10.1162/089976601750265045} for convergence results.
First, the values of $\Phi$ and $\Psi$ are initialized by using the prior PDFs of $Q$ and $R$
\begin{eqnarray}
\Phi^{(0)} &=& \frac{\int_{\mathcal Q} Q^{-1} \mathcal{W}^{-1} \left( Q; \bar{M}, \bar{m} \right) dQ }{\int_{\mathcal Q} \mathcal{W}^{-1} \left( Q; \bar{M}, \bar{m} \right) dQ} \label{eq:Phi0} \\
\Psi^{(0)} &=& \frac{\int_{\mathcal R} R^{-1} \mathcal{W}^{-1} \left( R; \bar{S}, \bar{s} \right) dR }{\int_{\mathcal R} \mathcal{W}^{-1} \left( R; \bar{S}, \bar{s} \right) dR} \, .  \label{eq:Psi0} 
\end{eqnarray}
Then, the algorithm of Table \ref{Fix-point-unbounded} is carried out.}

\begin{table}[tb]
\caption{ {Fixed point iteration for full-information VB estimation} }
\label{Fix-point-unbounded}
\mbox{} \\
\hrule \mbox{} \\
{\textbf{Set:}} 
$m = \bar{m}+t$, $s = \bar{s}+t$;  \\
\textbf{for} $k=1:N$ \\
\textbf{(a)} 
Compute $\hat{x}^{(k)}=\hat{x}\left(\Psi^{(k-1)},\Phi^{(k-1)}\right)$ via (\ref{x-mean});  \\
\textbf{(b)}
Compute ${P}^{(k)}=P\left(\Psi^{(k-1)},\Phi^{(k-1)}\right)$ via (\ref{x-P});  \\
\textbf{(c)}
Compute ${M}^{(k)}=M\left(\hat{x}^{(k)},P^{(k)}\right)$ via (\ref{Q-M});   \\
\textbf{(d)}
Compute ${S}^{(k)}=S\left(\hat{x}^{(k)},P^{(k)}\right)$ via (\ref{R-S});   \\
\textbf{(e)}
Compute: 
{
\begin{eqnarray}
 \quad \quad \Phi^{(k)} &=& \frac{\int_{\mathcal Q} Q^{-1} \mathcal{W}^{-1} \left( Q; {M}^{(k)}, {m} \right) dQ }{\int_{\mathcal Q} \mathcal{W}^{-1} \left( Q; {M}^{(k)}, {m} \right) dQ}  \label{eq:Phik} \\
\quad \quad \Psi^{(k)} &=& \frac{\int_{\mathcal R} R^{-1} \mathcal{W}^{-1} \left( R; {S}^{(k)}, {s} \right) dR }{\int_{\mathcal R} \mathcal{W}^{-1} 
\left( R; {S}^{(k)}, {s} \right) dR} \, ;  \label{eq:Psik}
\end{eqnarray}
}
\textbf{end} \\
\hrule
\end{table}

{
Notice that in the unconstrained case, i.e. $\mathcal Q = \mathbb S_+^{n_x}$
and { $\mathcal R = \mathbb S_+^{n_y}$,  }
the integrals in  (\ref{eq:Phik})-(\ref{eq:Psik}) can be computed in closed form.
In fact, in this case,
\begin{eqnarray}
\Phi^{(k)} &=& \int Q^{-1} \mathcal{W}^{-1} \left( Q; {M}^{(k)}, {m} \right) dQ = m \left [{M}^{(k)} \right ]^{-1}\\
\Psi^{(k)} &=& \int R^{-1} \mathcal{W}^{-1} \left( R; {S}^{(k)}, {s} \right) dR = s \, \left [{S}^{(k)} \right ]^{-1} \, .
\end{eqnarray}
Conversely, when $\mathcal Q \subset \mathbb S_+^{n_x}$ and $\mathcal R \subset \mathbb S_+^{n_y}$,
the integrals in   (\ref{eq:Phik})-(\ref{eq:Psik})
can no longer be computed in closed form  
but can, anyway, be easily approximated to any desired accuracy via Monte Carlo integration. Analogous considerations hold
for the integrals in  (\ref{eq:Phi0})-(\ref{eq:Psi0}). The discussion on the application of Monte Carlo integration in the considered framework 
is deferred
to the next section.
}

%
%

Notice that the above inference, relying on the whole measurement sequence $y_{1:t}$ up to time $t$,
is characterized by memory and computational complexity growing with time.
For the sake of implementation, a \textit{moving horizon}  approximation of finite fixed length $T \geq 1$
will be considered hereafter, by only exploiting at time $t$ the measument sub-sequence $y_{t-T+1:t}$ in order to estimate
the state sub-trajectory $\hat{x}_{t-T:t}$.

\section{Variational Bayes Moving-horizon estimation algorithm}
\label{bounded-MHE}

The purpose of this  section is to  make the proposed approach  recursive by means of the MHE approximation, 
where the estimation results at the current time index is used as the initial value for the next moving horizon estimation.

Specifically, suppose that the information collected up to time $t-T$ can be approximately summarized by the PDF 
\begin{eqnarray}
&&p\left( x_{t-T}, Q, R \,|\, y_{1:t-T}\right)  \nonumber \\
&&\propto  \mathcal{N} \left( x_{t-T}; \bar{x}_{t-T}, \bar{P}_{t-T} \right)  \,
\mathcal{W}^{-1} \left( Q; \bar{M}_{t-T}, \bar{m}_{t-T} \right)   \nonumber \\
&&\quad \times 
 \mathcal{W}^{-1} \left( R; \bar{S}_{t-T}, \bar{s}_{t-T} \right)  \, \mathbf{1}_{\mathcal{Q}} (Q)  \, \mathbf{1}_{\mathcal{R}} (R) \, .
\label{prior-bounded}
\end{eqnarray}
Then, we can apply the previously outlined VB approach to compute an approximation of the form
\begin{eqnarray}
&&p\left( x_{t-T:t}, Q, R \,|\, y_{t-T+1:t}\right)  \nonumber \\
&&\propto  \mathcal{N} \left( x_{t-T:t}; \hat{x}_{t-T:t|T}, P_{t-T:t|T} \right)   \nonumber \\
&&\hspace{.22cm} \times \mathcal{W}^{-1} \left( Q; M_{t}, m_{t} \right)  \, 
\mathcal{W}^{-1} \left( R; S_{t}, s_{t} \right)  \,  \mathbf{1}_{\mathcal{Q}} (Q) \, \mathbf{1}_{\mathcal{R}} (R)
\label{approxi-bounded}
\end{eqnarray}
given the measurement sequence $y_{t-T+1:t}$ and the prior knowledge summarized by 
$\bar{x}_{t-T}$, $\bar{P}_{t-T}$, $\bar{M}_{t-T}$, $\bar{m}_{t-T}$, $\bar{S}_{t-T}$ and $\bar{s}_{t-T}$, 
which are initialized at time $t=T$ from the prior distributions of $x_0$, $Q$, $R$.
Notice that, in order to account for the moving horizon, instead of the function $\omega (\Psi, \Phi)$ of Proposition 1
we consider
\begin{equation}
\omega_{t} \left(\Psi \right) = 
\begin{bmatrix} C' \Psi y_t   \\    \vdots    \\   C' \Psi y_{t-T+1}   \\   
\bar{P}_{t-T}^{-1} \, \bar{x}_{t-T} 
\end{bmatrix}
\label{omegat} \, .
\end{equation}
Similarly, instead of the matrix $\Omega (\Psi, \Phi)$ in (\ref{Omega}), we consider the matrix $\Omega_t (\Psi, \Phi)$ defined 
with respect to the sliding window
$[t-T,t]$ with $\bar P_0$ replaced by $\bar P_{t-T}$.  
As to Proposition \ref{proposition-2}, instead of the functions $M(\hat x,P)$ and $S (\hat x, P)$, we consider
\begin{eqnarray}
&M_t&\left( \hat{x}_{t-T:t},P_{t-T:t}\right) \nonumber \\ && = \bar{M}_{t-T} + \sum_{i=t-T}^{t} \Big[ \left(\hat{x}_i-A\hat{x}_{i-1}\right) \left(\hat{x}_i-A\hat{x}_{i-1}\right)'    \nonumber \\
&& \quad\qquad+  P_{i}  + AP_{i-1}A' - P_{i,i-1}A'  - AP_{i-1,i}  \Big]  
\label{Q-Mt} \\
&S_t&\left(\hat{x}_{t-T:t},P_{t-T:t}\right) = \bar{S}_{t-T} \nonumber  \\ && + \sum_{i=t-T}^{t} \Big[ \left(y_i-C\hat{x}_{i}\right) 
\left(y_i- C\hat{x}_{i}\right)'  
+ C P_{i} C'   \Big] \, .
\label{R-St}
\end{eqnarray}

In order to go from time $t$ to time $t+1$, one iteration of the KF can be performed 
starting from the most recent estimates $\hat{Q}_t$ and $\hat{R}_t$ of the PNCM and MNCM.
The latter can be computed via the integrals
{
\begin{eqnarray}
\hat Q_t &=& \frac{\int_{\mathcal Q} Q \, \mathcal{W}^{-1} \left( Q; {M}_t, {m_t} \right) dQ }{\int_{\mathcal Q} \mathcal{W}^{-1} \left( Q; {M}_t, {m_t} \right) dQ}  \label{integ-Qt-bounded} \\
\hat R_t &=& \frac{\int_{\mathcal R} R \,  \mathcal{W}^{-1} \left( R; {S}_t, {s_t} \right) dR }{\int_{\mathcal R} \mathcal{W}^{-1} 
\left( R; {S}_t, {s_t} \right) dR} \, .  \label{integ-Rt-bounded}
\end{eqnarray}
}

Then, we can update $\bar{x}_{t-T+1}$ and $\bar{P}_{t-T+1}$  
as follows
\begin{eqnarray}
\tilde{x}_{t-T+1} &=& A \bar{x}_{t-T} \label{x_prediction} \\
\tilde{P}_{t-T+1}  &=& A \bar{P}_{t-T} A' + \hat{Q}_{t}  \label{P_prediction}  \\
K_{t-T+1} &=& \tilde{P}_{t-T+1}  C' \left ( C   \tilde{P}_{t-T+1}  C' + \hat R_t \right )^{-1} \\
\bar{P}_{t-T+1}  &=&  \left ( I - K_{t-T+1}  C \right ) \tilde{P}_{t-T+1}  \label{P_propogation} \\
\bar{x}_{t-T+1} &=& \tilde{x}_{t-T+1}  + K_{t-T+1}  \left (  y_{t-T+1} - C  \tilde{x}_{t-T+1}  \right )  .
\label{x_propogation}
\end{eqnarray}

For the time propagation of  the parameters $\bar{M}_{t-T}$ and $\bar{m}_{t-T}$, 
following \citep{8025799},
we can set 
\begin{eqnarray} 
\bar{M}_{t-T+1} &=& \rho \, M_t   \label{pre-M}    \\
\bar{m}_{t-T+1} &=& \rho \, \left(m_t-n_x-1\right)+n_x+1    \label{pre-m}
\end{eqnarray}
where $\rho \in (0,1)$ denotes a forgetting factor.
Similarly, $\bar{S}_{t-T+1}$ and  $\bar{s}_{t-T+1}$  can be obtained by  
\begin{eqnarray} 
\bar{S}_{t-T+1} &=& \rho \, S_t  \label{pre-S}    \\
\bar{s}_{t-T+1} &=& \rho \, \left(s_t-n_y-1\right)+n_y+1.     \label{pre-s}
\end{eqnarray}

To summarize, the implementation of  the proposed VB MHE algorithm
is outlined in Table \ref{alg:Algorithm 2}.

\begin{table}  
\caption{{VB-MHE algorithm}  }
\label{alg:Algorithm 2}
{
\mbox{} \\
\hrule \mbox{} \\
\textbf{Inputs:}
$y_{t-T+1:t}$, $\bar{x}_{t-T}$, $\bar{P}_{t-T}$, 
$\bar{M}_{t-T}$, $\bar{m}_{t-T}$, 
$\bar{S}_{t-T}$, $\bar{s}_{t-T}$, ${\Phi}_{t-1}$, ${\Psi}_{t-1}$ \\   
\hrule \mbox{} \\
\textbf{(1)} {\bf Variational iterations:} \\
Set $\Phi_{t}^{(0)}=\Phi_{t-1}$,  $\Psi_{t}^{(0)}=\Psi_{t-1}$ ; \\
Set $m_t = \bar m_{t-T} +T$, $s_t = \bar s_{t-T} + T$; \\
\textbf{for} $k=1:N$   \\
\textbf{(a)} Compute $P_{t-T:t}^{(k)} =  \Omega^{-1}_t \left ( \Psi_t^{(k-1)}, \Phi_t^{(k-1)} \right ) $; \\  
\textbf{(b)} Compute   
$ \hat x^{(k)}_{t-T:t} = P_{t-T:t}^{(k)} \, \, \omega_t \left ( \Psi_t^{(k-1)} \right ) $; \\
\textbf{(c)} Compute ${M}_t^{(k)} = M_t\left(\hat{x}_{t-T:t}^{(k)},P_{t-T:t}^{(k)}\right)$; \\
\textbf{(d)} Compute ${S}_t^{(k)} = S_t\left(\hat{x}_{t-T:t}^{(k)},P_{t-T:t}^{(k)}\right)$;  \\
 \textbf{(e)} 
 Compute \begin{eqnarray}
 \quad \quad \Phi^{(k)}_t &=& \frac{\int_{\mathcal Q} Q^{-1} \mathcal{W}^{-1} \left( Q; {M}^{(k)}_t, {m}_t \right) dQ }{\int_{\mathcal Q} \mathcal{W}^{-1} \left( Q; {M}^{(k)}_t, {m}_t \right) dQ}  \label{eq:Phikt} \\
\quad \quad \Psi^{(k)}_t &=& \frac{\int_{\mathcal R} R^{-1} \mathcal{W}^{-1} \left( R; {S}^{(k)}_t, {s}_t \right) dR }{\int_{\mathcal R} \mathcal{W}^{-1} 
\left( R; {S}^{(k)}_t, {s}_t \right) dR} \, ;  \label{eq:Psikt}
\end{eqnarray}

\textbf{end for}   \\
Set 
$\hat{x}_{t-T:t|t} = \hat{x}_{t-T:t}^{(N)}$,  $P_{t-T:t|t} = P_{t-T:t}^{(N)}$; \\
Set ${M}_t = M_t^{(N)}$,  $S_t = S_t^{(N)}$; \\
Set $\Phi_t = \Phi_t^{(N)}$, $\Psi_t = \Psi_t^{(N)}$; \\
Compute $\hat{Q}_t $ and  $\hat{R}_t $ via  (\ref{integ-Qt-bounded})-(\ref{integ-Rt-bounded}); \\

\textbf{(2)} {\bf Time update for moving horizon:} \\
\textbf{(a)} Compute $\bar{M}_{t-T+1}$, $\bar{m}_{t-T+1}$, $\bar{S}_{t-T+1}$, $\bar{s}_{t-T+1}$ 
for the next moving horizon filtering recursion via (\ref{pre-M})-(\ref{pre-s});

\textbf{(b)} Compute $\bar{x}_{t-T+1}$ and $\bar{P}_{t-T+1}$ 
via KF with estimated $\hat{Q}_t$ and $\hat{R}_t$ via (\ref{x_prediction})-(\ref{x_propogation}); 
 \\   
\hrule \mbox{} \\
\textbf{Outputs:} $\hat{x}_{t|t}$, $P_{t|t}$, $\hat{Q}_t$, $\hat{R}_t$, $\Phi_t$, $\Psi_t$  
$\bar{x}_{t-T+1}$, $\bar{P}_{t-T+1}$, 
$\bar{M}_{t-T+1}$, $\bar{m}_{t-T+1}$, 
$\bar{S}_{t-T+1}$, $\bar{s}_{t-T+1}$ \\
\hrule
}
\end{table}  

{ The integrals in  (\ref{integ-Qt-bounded})-(\ref{integ-Rt-bounded}) can be approximated to any desired degree of accuracy
by means of Monte Carlo integration with importance sampling.
This amounts to drawing $J$ samples $Q_j$ and $R_j$,with  $j=1,2, \ldots,J$, from suitable proposal distributions
$\pi_Q (Q)$ and $\pi_R (R)$, respectively, and then setting
\begin{eqnarray} 
&&\hat{Q}_{t} = \frac
{ \sum\limits_{j=1}^{J} Q_{j}  \, \frac{\mathcal W^{-1} \left( Q_j; M_t, m_t \right)}{\pi_Q \left(Q_j\right)} \, 
\mathbf{1}_{\mathcal{Q}}\left(Q_j\right) } 
{ \sum\limits_{j=1}^{J}   \, \frac{\mathcal W^{-1} \left( Q_j; M_t, m_t \right)}{\pi_Q \left(Q_j\right)} \, 
\mathbf{1}_{\mathcal{Q}}\left(Q_j\right) }
\label{samp-Qt-bounded}  \\
&&\hat{R}_{t} = \frac
{ \sum\limits_{j=1}^{J} R_{j}  \, \frac{\mathcal W^{-1} \left( R_j; S_t, s_t \right)}{\pi_R \left(R_j\right)} \, 
\mathbf{1}_{\mathcal{R}}\left(R_j\right) } 
{ \sum\limits_{j=1}^{J}   \, \frac{\mathcal W^{-1} \left( R_j; S_t, s_t \right)}{\pi_R \left(R_j\right)} \, 
\mathbf{1}_{\mathcal{R}}\left(R_j\right) } \, .
\label{samp-Rt-bounded} 
\end{eqnarray}
A reasonable choice for the proposal distributions amounts to setting
\begin{eqnarray}
\pi_Q\left(Q\right) &=& \mathcal{W}^{-1} \left(Q; \left(m_t-n_x-1\right) \hat{Q}_{t-1}, m_t \right) \\
\pi_R\left(R\right) &=& \mathcal{W}^{-1} \left(R; \left(s_t-n_y-1\right) \hat{R}_{t-1}, s_t \right) \, .
\end{eqnarray}
In fact, since the previous estimates $\hat{Q}_{t-1}$ and  $\hat{R}_{t-1}$ belong by construction to $\mathcal Q$ and $\mathcal R$, respectively,
then the above choice ensures that most of the samples are drawn inside those sets. 
The integrals in (\ref{eq:Phikt}) and (\ref{eq:Psikt}) can be computed  a similar way.
}

As will be shown in the next section, the resulting recursive estimation algorithm, under the assumption of bounded $\mathcal{Q}$ and $\mathcal{R}$,
turns out to be mean-square stable for any $J$ (number of samples), any $N$ (number of VB iterations), 
and any $T$ (length of the moving horizon window).



\section{{Stability analysis} }
\label{stability}
The stability of the proposed VB MHE adaptive filter of Algorithm \ref{alg:Algorithm 2} 
is analyzed in this section in terms of boundedness of the estimation error ${e}_{t}=x_{t} - \hat{x}_{t|t}$.
{To this end,
the following assumptions are needed. 

\textbf{A1.} The pair $(A,C)$ is detectable.

\textbf{A2.} There exist $\underline{\alpha}$ and $\overline{\alpha}$ with 
$0 < \underline{\alpha} \le \overline{\alpha}$ such that  $\underline{\alpha} I \leq Q \leq \bar{\alpha} I$, 
$\forall \, Q \in \mathcal{Q}$.

\textbf{A3.} There exist $\underline{\beta}$ and $\overline{\beta}$ with 
$0 < \underline{\beta} \le \overline{\beta}$ such that  $\underline{\beta} I \leq R \leq \bar{\beta} I$, 
$\forall \, R \in \mathcal{R}$.
}

Under the stated assumptions, the following result descends from classical results on KF  \citep{jazwinskiStotic}.

\begin{lem}
\label{lemma-2}  
Let assumptions A1-A3 be satisfied.
Then, there exist real numbers $\underline{p}$ and $\overline{p}$ 
{ with $0 < \underline{p} \le \overline{p} $} such that
\begin{eqnarray} 
\underline{p} I \leq \bar{P}_{t-T+1} \leq \overline{p} I 
\end{eqnarray}
for any $t \ge T$.
\end{lem}
{ {\bf Proof.\,} 
 Notice that $\bar{P}_{t-T+1}$  is the covariance matrix of a KF constructed by using the estimated covariance matrices
 $\hat Q_t$ and $\hat R_t$ in place of the true ones. By construction, the estimates $\hat Q_t$ 
 and $\hat R_t$ belong to the sets $\mathcal Q$ and $\mathcal R$ irrespectively of the number of samples $J$,
 the number of VB iterations $N$ and the window length $T$. Hence,  $\underline{\alpha} I \leq \hat Q_t \leq \bar{\alpha} I$
 and $\underline{\beta} I \leq \hat  R_t \leq \bar{\beta} I$ for any $t \ge T$. Then, the existence of uniform upper and lower bounds
  for $\bar{P}_{t-T+1}$ 
 can be proved as in  \citep{jazwinskiStotic}. \hfill $\square$
}

Let us now recall the following result.

\begin{lem}
\label{lemma-1} 
If a stochastic process $V_t\left(\varepsilon_t\right)$ satisfies the following conditions 
(where $\underline{\gamma} \,$, $\overline{\gamma} \,$, $\lambda$ and $\mu$ are real numbers satisfying 
{ $0 < \underline{\gamma} \le \overline{\gamma}$,} $0\le \lambda < 1$ and $\mu>0$): 
\begin{eqnarray} 
&&\underline{\gamma} \, \Vert \varepsilon_t \Vert^2  \leq  V_t\left(\varepsilon_t\right)  \leq    \bar{\gamma} \, \Vert \varepsilon_t \Vert^2  
\label{lyapunov1} \\
&& \left \{ E \left[ V_t \left( \varepsilon_t \right) \right] \right \}^{1/2} \le \lambda \left \{ E \left[  V_{t-1} \left( \varepsilon_{t-1} \right) \right] \right \}^{1/2} +  \mu  \label{lyapunov2}
\end{eqnarray}
then the stochastic process is exponentially bounded in mean square, i.e
\begin{eqnarray} 
E \left[ \Vert \varepsilon_t \Vert^2 \right]   \leq 
&&  \frac{2 \, \overline{\gamma} \,}{\underline{\gamma} \,}  E \left[ \Vert \varepsilon_0 \Vert^2 \right]  \lambda^{2t}  + \frac{2}{\underline{\gamma} \,} \left (  \mu \sum_{i=0}^{t-1} \lambda^i \right )^2 \, .
\end{eqnarray}
\end{lem}

Based on Lemma \ref{lemma-2} and Lemma \ref{lemma-1},  the following stability result can be proven.

\begin{thm}
\label{theorem-1}
Let assumptions A1-A3 be satisfied.
%
%
%
%
Then  the state estimation error sequence 
 ${e}_{t}=x_{t} - \hat{x}_{t |t} $ is uniformly bounded in mean square.
\end{thm}

{\bf{Proof.\,}}
We first prove that $\bar{e}_{t} = x_{t} - \bar x_{t} $ is {{uniformly}} bounded 
and then prove that ${e}_{t}$ is uniformly bounded as well.

Let us first define a candidate Lyapunov function
\begin{eqnarray} 
V_{t+1} (\tilde{e}_{t+1}) = \tilde{e}_{t+1}'  \tilde{P}_{t+1}^{-1}  \tilde{e}_{t+1}
\end{eqnarray}
where $\tilde{e}_{t+1} = x_{t+1} -  \tilde x_{t+1} $. 
In view of (\ref{P_prediction}) and Lemma  \ref{lemma-2}, we have
 $\underline{\alpha} I   \leq   \tilde{P}_{t+1}  \leq  \overline{p} \, \| A \|^2 \, I + \overline{\alpha} I $.
Hence, the Lyapunov candidate function satisfies (\ref{lyapunov1}) with
$\overline{\gamma} = 1/ \underline{\alpha} $ and $\underline{\gamma} = 1/ (\overline{p} \, \| A \|^2 + \overline \alpha)$.

Next, in view of (\ref{x_prediction})-(\ref{x_propogation}), we can write
\begin{eqnarray} 
\tilde e_{t+1} = A \, (I - K_{t} C) \, \tilde e_{t} + A \,  K_{t} \, v_{t} + w_{t} \, .
\end{eqnarray}
Consider now the square root expected value $\tilde V_{t+1} (e) = \left \{ E \left [ V_{t+1} (e) \right ]\right \}^{1/2}$ 
of the candidate Lyapunov function.
Since $\tilde V_{t+1} (e) $ is a norm, we can apply the triangular inequality and write
\begin{eqnarray}
\tilde V_{t+1} (e_{t+1})  
&\le& \tilde V_{t+1} (A \, (I - K_{t} C) \, \tilde e_{t}) \nonumber \\ && + \tilde V_{t+1} (A \,  K_{t} \, v_{t} ) +  
\tilde V_{t+1} (w_{t} )  \, . \label{condition-2}
\end{eqnarray}
Notice that 
\begin{eqnarray}
\tilde V_{t+1} (w_{t} ) \le (\overline \gamma)^{1/2} \left \{ E \left [ \| w_t \|^2 \right ] \right \}^{1/2} \le [ \overline \gamma \,  {\rm tr} \, (Q) ]^{1/2} \, .
\end{eqnarray}
Further, under the stated assumptions, the Kalman gain is bounded in that $\| K_{t-T} \| \le \| \tilde{P}_{t-T} \|  \, \| C \| \, \underline \beta^{-1} \le (\overline{p} \, \| A \|^2 + \overline \alpha) \| C \| \, \underline \beta^{-1}  $. Then, we have
\begin{eqnarray}
\tilde V_{t+1} (A \,  K_{t} \, v_{t} ) &\le& (\overline \gamma)^{1/2} \left \{ E \left [ \| A \,  K_{t} \, v_{t}  \|^2 \right ] \right \}^{1/2}  \nonumber \\
&\le&  [ \overline \gamma \,  {\rm tr} \, (R) ]^{1/2} { \mathcal{K} } 
\| A \| 
\end{eqnarray}
where $\mathcal{K} = (\overline{p} \, \| A \|^2 + \overline \alpha) \| C \| \, \underline \beta^{-1}  $.
Finally, with standard manipulations (see \citep{7403006}),  under the stated assumptions it can be shown that
\begin{eqnarray}
V_{t+1} (A \, (I - K_{t} C) \, \tilde e_{t}) \le \lambda^2 V_{t} (\tilde e_{t})
\end{eqnarray}
for some  $\lambda$ with $0\le \lambda < 1$. Hence, from (\ref{condition-2}), we can derive  (\ref{lyapunov2}) 
by setting $\mu = [ \overline \gamma \,  {\rm tr} \, (Q) ]^{1/2}  +  [ \overline \gamma \,  {\rm tr} \, (R) ]^{1/2} { \mathcal{K} }   \| A \| $.
%
%
Then,
according to Lemma \ref{lemma-1}, 
$\tilde{e}_{t}$ is mean-square bounded  under the  
given assumptions.


Further, since
\begin{equation}
\bar{e}_{t} = (I - K_{t} C) \, \tilde e_{t} + K_{t} \, v_{t} \, ,
\end{equation}
we have that 
\begin{eqnarray}
E\left[ \Vert \bar{e}_{t} \Vert^2 \right]   \leq   
2 (1 + \kappa \| C \|)^2 E\left[ \Vert \tilde {e}_{t} \Vert^2 \right]  + 2 {\mathcal{K} } ^2 \, 
{\rm tr(R)} \, .
\label{eq:bare}
\end{eqnarray}
Therefore, the estimation error $\bar{e}_{t}$ is also bounded in the mean-square sense. 

Next, 
we will prove that  ${e}_{t-T:t}$
is bounded. 
To this end, it is convenient to decompose $\hat \Omega_t = \Omega_t (\Psi_t^{(N-1)}, \Phi_t^{(N-1)})$
as follows
\begin{eqnarray}
\hat \Omega_t &=& \Omega_{t,1} + \Omega_{t,2} + \Omega_{t,3}
\label{Ome-stability}
\end{eqnarray}
where
\begin{eqnarray}
\Omega_{t,1} &=& 
\begin{bmatrix}  
&C'\Psi_t^{(N-1)} C   &0  &\cdots &  &    0   \\
&0   &C'\Psi_t^{(N-1)} C     &   &  &\vdots     \\
&\vdots  &   &\ddots  &  &  \\
&    &   &  &C'\Psi_t^{(N-1)} C   & 0   \\
&  0  &  \cdots &    &   0 &0    \end{bmatrix}  \nonumber \\
\Omega_{t,2} &=& \begin{bmatrix}  
0   &  \cdots        &  0  & 0     \\
 \vdots    &\ddots          &   &  \vdots    \\
  0   &           &   0 &  0 \\
   0 &  \cdots          &  0  &    \bar{P}^{-1}_{t-T}   \end{bmatrix}    \nonumber \\
\Omega_{t,3} &=&  \begin{bmatrix}  
&\Omega_{3}^{-}  &\Omega_{3}^{\circ}  &0   &\cdots  &0   \\
&\Omega_{3}^{\times}  &\; \Omega_{3}^{-} + \Omega_{3}^{+}  \;  &\Omega_{3}^{\circ}  & &\vdots   \\
&0 &\Omega_{3}^{\times}  &\ddots  &\ddots  &\\
&\vdots  &   &\ddots  &\; \Omega_{3}^{-} + \Omega_{3}^{+}  \; &\Omega_{3}^{\circ}  \\
&0   &\cdots  & &\Omega_{3}^{\times}   &\Omega_{3}^{+}     \end{bmatrix}    \label{Omega_3}
\end{eqnarray}
with 
\begin{eqnarray} 
\begin{cases} 
\Omega_{3}^{-} &= \quad \Phi^{(N-1)}  \\
\Omega_{3}^{+} &= \quad A^\mathrm{T} \Phi^{(N-1)}  A   \\
\Omega_{3}^{\times} &= \quad -A' \Phi^{(N-1)}   \\
\Omega_{3}^{\circ}  &= \quad -\Phi^{(N-1)}  A   \, . 
\end{cases}
\label{Omega_3_para}
\end{eqnarray}
Further, it is an easy matter to check that $\hat \omega_t = \omega_t (\Psi_t^{(N-1)})$ can be decomposed as follows
\begin{eqnarray}
\hat \omega_t  &=& \omega_{t,1} + \omega_{t,2} + \omega_{t,3} 
\label{ome-stability} 
\end{eqnarray}
where
\begin{eqnarray}
\omega_{t,1} &=& \Omega_{t,1} x_{t-T:t} 
\nonumber \\
\omega_{t,2} &=& \begin{bmatrix} 0 \\ \vdots \\ 0 \\ \bar P_{t-T} \bar {x}_{t-T} \end{bmatrix} \; \omega_{t,3} = \begin{bmatrix} 
& C'\Psi_t^{(N-1)}  v_t   \\   &\vdots    \\ &    \\  &   \\  & C'\Psi_t^{(N-1)} v_{t-T +1} \\ & 0  
\end{bmatrix} \, .     \label{omega_3}
\end{eqnarray}

Clearly, from   (\ref{Ome-stability}), the true state trajectory satisfies the identity  
\begin{eqnarray}
x_{t-T:t} = \hat \Omega_t^{-1} \left ( \Omega_{1,t} + \Omega_{2,t}  + \Omega_{3,t}  \right ) x_{t-T:t} \, .
\label{true-x-stability}
\end{eqnarray}
Further, for the estimated state trajectory we have
\begin{eqnarray}
\hat x_{t-T:t|T} &=& \hat \Omega_t^{-1}  \hat \omega_t = \hat \Omega_t^{-1}  \left ( \omega_{t,1} + \omega_{t,2} + \omega_{t,3}\right ) \nonumber \\
&=&   \hat \Omega_t^{-1}   \Omega_{t,1} x_{t-T:t} + \hat \Omega_t^{-1} ( \omega_{t,2} + \omega_{t,3} )  \, . \label{estimate-x-stability}
\end{eqnarray}
Hence, by subtracting the two latter equations, we get
\begin{eqnarray}
\lefteqn{x_{t-T:t}  - \hat x_{t-T:t|T} } \nonumber \\
&& =  \hat \Omega_t^{-1} \left ( \Omega_{2,t}  x_{t-T:t} - \omega_{2,t} +   \Omega_{3,t}  x_{t-T:t} - \omega_{3,t}  \right ) \, .  \label{error-x-stability}
\end{eqnarray}
Notice that
\begin{eqnarray}
 \Omega_{2,t} \,  x_{t-T:t} - \omega_{2,t}  = \begin{bmatrix} 0 \\ \vdots \\ 0 \\ \bar P_{t-T}^{-1} \, \bar e_{t-T} \end{bmatrix}
 \label{omega2}
\end{eqnarray}
which turns out to be uniformly bounded in mean square as previously shown. 
Notice also that
\begin{eqnarray}
 \Omega_{3,t}  x_{t-T:t}  = \begin{bmatrix}  
& \Phi^{(N-1)}_t w_{t-1}     \\
&-A'  \Phi^{(N-1)}_t w_{t-1} +  \Phi^{(N-1)}_t w_{t-2} \\
& \vdots  \\
&-A'  \Phi^{(N-1)}_t w_{t-T+1} +   \Phi^{(N-1)}_t w_{t-T}   \\
&-A'  \Phi^{(N-1)}_t w_{t-T}   \end{bmatrix} .
\label{Omega3}
\end{eqnarray}
Notice finally that,  irrespectively of the number of samples $J$,
 the number of VB iterations $N$ and the window length $T$, by construction the matrices
 $\Phi^{(k)}_t$ and $\Psi^{(k)}_t$ can be bounded as 
$\overline{\alpha}^{-1} I \leq \hat \Phi^{(k)}_t \leq \underline{\alpha}^{-1} I$
 and $\overline{\beta}^{-1} I \leq \hat  \Psi^{(k)}_t \leq \underline{\beta}^{-1} I$ for any $t \ge T$. 
 As a consequence, all the matrices involved in (\ref{error-x-stability}) are uniformly bounded.
 Then, in view of (\ref{omega_3}) (\ref{error-x-stability}),(\ref{omega2}), (\ref{Omega3}), we can conclude that
 there exist suitable constants $c_1$, $c_2$ and $c_3$ such that
 \begin{eqnarray}
 E \left [ \| e_t \|^2 \right ] \le c_1  E \left [ \| \bar e_{t-T} \|^2 \right ] + c_2 {\rm tr} (Q) + c_3 {\rm tr} (R) \, .
 \end{eqnarray}
By combining the latter inequality with (\ref{eq:bare}), the uniform mean square boundedness of $e_t$ follows.
\hfill $\square$

\begin{remark}
The proposed algorithm has been developed under the assumption that the unknown PNCM and MNCM are nearly constant within the sliding window
$[t-T,t]$. This condition is satisfied whenever the unknown covariance matrices are constant or their variations are slow compared to the size $T$ of the sliding window. The forgetting factor $\rho$ in the time propagation (\ref{pre-M})-(\ref{pre-s}) can be tuned
so as to make the filter more able to promptly detect variations in the unknown covariance matrices (by choosing $\rho$ close to $0$) or  
to improve the estimation accuracy for nearly constant matrices (by choosing $\rho$ close to $1$). Nevertheless,
the stability of the estimation error is guaranteed for any choice of $\rho$ and for any time-varying $Q_t$ and $R_t$ provided that they remain uniformly bounded.
\end{remark}

\section{Simulations}
\label{simulation} 

To assess the performance of the proposed adaptive VB MHE filter, 
a $2$-dimensional target tracking example is considered in this section.
The target moves according to (\ref{state-eq}) 
with state $x=\left[\xi_t, \eta_t, \dot{\xi}_t, \dot{\eta}_t \right]'$,
where $\left(\xi_t, \eta_t\right)$ and $(\dot{\xi}_t, \dot{\eta}_t )$ 
denote target position and velocity in Cartesian coordinates, respectively. The 
state transition matrix is
$A = \left[\begin{array}{cc} 
I_{2} & \quad \mathscr{T} I_{2} \\  
0 & \quad I_{2} \end{array}\right]$
where $I_n$ is the the $n \times n$ identity and $\mathscr{T}=1 \, [s]$ the sampling interval. 
The target position coordinates are measured according to
the measurement model (\ref{measurement-eq}) 
with $C = \left[I_{2}  \quad 0 \right]$.
The unknown process and measurement noise covariances
are supposed to belong to the bounded sets
\begin{eqnarray}
\mathcal Q &=& \{ Q \in {\mathbb S}_+^4 : \, 0.001\,Q_0 \leq Q \leq 1000\,Q_0\} \\
\mathcal R &=& \{ R \in {\mathbb S}_+^2 : \, 0.1\,R_0 \leq R \leq 10\,R_0\} 
\end{eqnarray}
where the nominal PNCM and MNCM are given by \\
\[Q_0 = \left[\begin{array}{cccc} 
1/3 &0 &1/2  &0  \\  
0 & 1/3  &0 &1/2 \\
1/2  &0  &1  &0 \\
0 &1/2  &0  &1
\end{array}\right], \quad R_0 =  100 \, \left[\begin{array}{cc} 
1 &0.5 \\  
0.5 &1 \end{array}\right] \, . \]

Monte Carlo simulations with $50$ independent trials of duration $t=500\, [s]$  have been carried out  
to compare the nominal Kalman filter (NKF), the conventional VB filter of \citet{8025799}
and the sliding window variational Kalman filter (VB Sliding Window, VB SW) of \citet{9096408} 
with the proposed adaptive VB MHE filter.
Initial state and covariance for all filters are set to: 
$x_0 = \left[ \, 0 \left[m\right], 10 \left[m\right], 0 \left[m/s\right], 10 \left[m/s\right] \,  \right]'$ and  
$P_0 = {\rm diag} \left\lbrace 100[m^2], 100[m^2], 100[m^2/s^2], 100[m^2/s^2] \right\rbrace$.
For the nominal KF, the PNCM and MNCM are set to $Q_0$ and $R_0$;
for the conventional VB, the nominal PNCM is set to $Q_0$ while the PECM and MNCM are estimated adaptively;
the VB parameters for the conventional VB, VB Sliding Window and proposed VB MHE 
are set as in \citep{8025799}, i.e., 
 $\rho=0.9$,
 $\hat{S}_{0|0}^i= \kappa \,R_0$, $\hat{s}_{0|0}^i = \kappa + n_y +1$, $\kappa=3$, 
 $\tau=3$, $N=1$.
Further, for the VB Sliding Window and VB MHE, 
different values of the window length $T$ are considered, i.e. 
$T \in \{ 4, 5, 10, 20 \}$.
The number of importance samples of the proposed bounded VB MHE is set to $J=100$. 
The unknown true PNCM and MNCM are set to 
{ $Q=50\,Q_0$}
and  $R=3 \,R_0$, respectively.

\begin{figure} [tbp]
\centering
	\begin{minipage}[tbp]{0.5\textwidth}  
    \subfloat[] {   \includegraphics[width=1\textwidth]{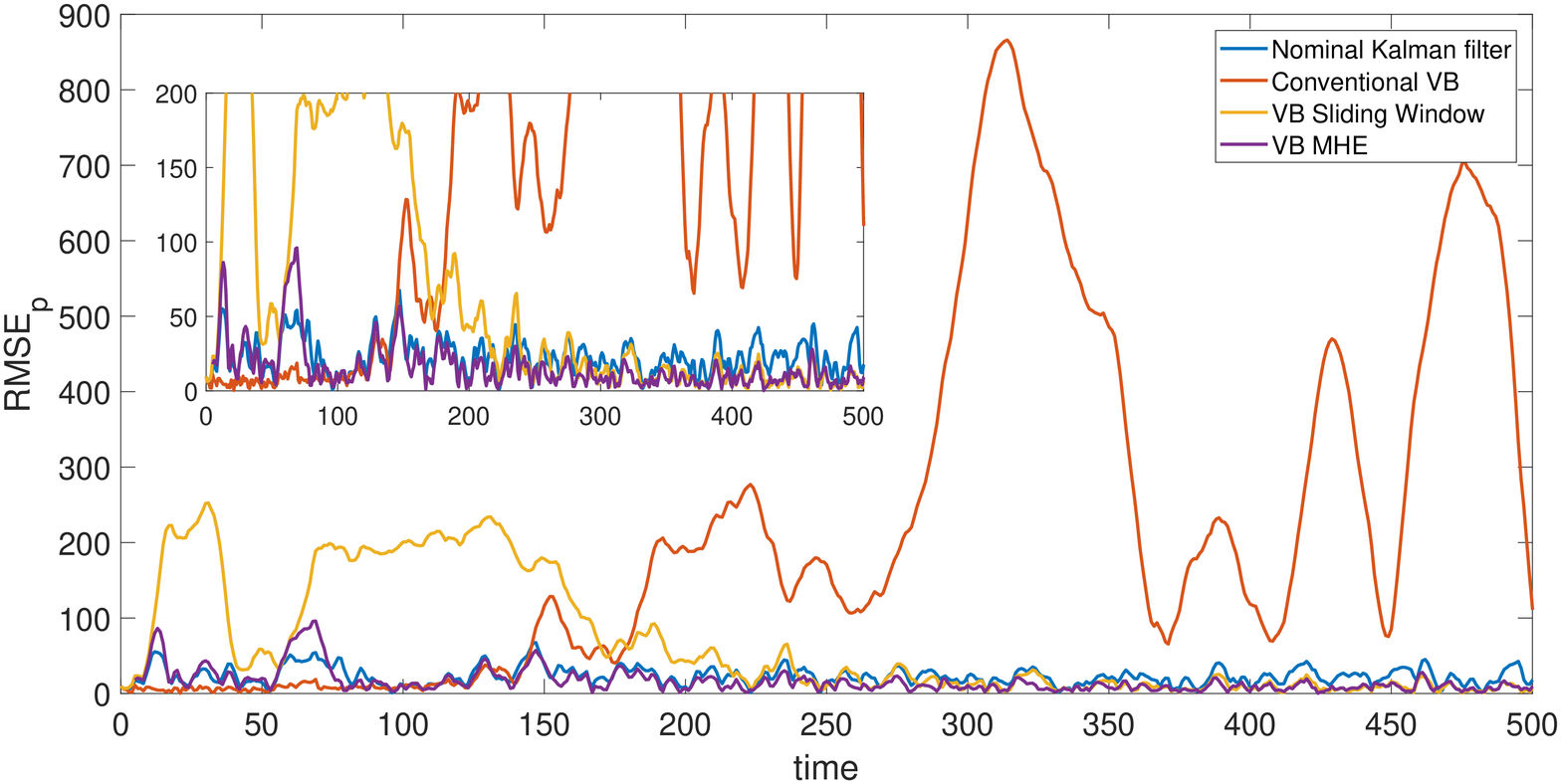}   } 
    \label{RMSE_p_L5}  
    \end{minipage} \\   
    \begin{minipage}[tbp]{0.5\textwidth}    
    \subfloat[] {   \includegraphics[width=1\textwidth]{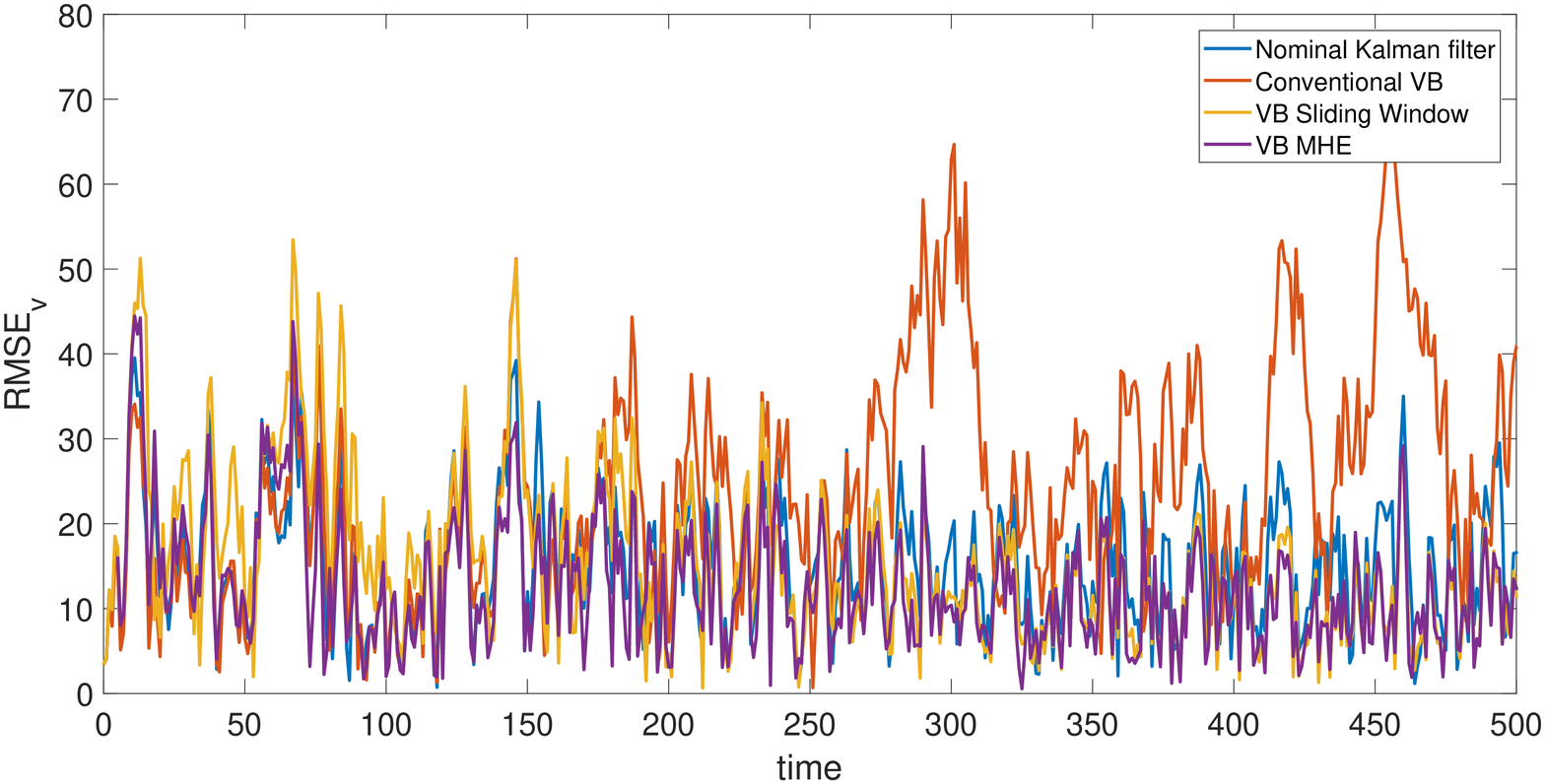}   }  
    \label{RMSE_v_L5}  
    \end{minipage}   
\caption{Position (a) and velocity (b) RMSEs ($T=5$)}
\label{RMSE_L5}
\end{figure}

For the filtering performance assessment, the
\textit{root mean square error} (RMSE) versus time and the time-\textit{averaged RMSE} (ARMSE) 
for position and velocity over  the  whole simulation  
are provided in Fig. 
\ref {RMSE_L5},
and respectively Tables \ref{ARMSE-p}-\ref{ARMSE-v}, 
demonstrating the outperformance of the proposed filter with respect to the others. 
It can be seen from Tables \ref{ARMSE-p}-\ref{ARMSE-v} that, 
when $T=20$, the proposed VB MHE filter provides improvement with respect to the conventional VB, nominal KF, VB Sliding Window of $96\% $, $59\%, 42\%$ in position ARMSE
and $56\% $, $29\%$, $11\%$ in velocity ARMSE. 
Conversely, when $T=4$, the corresponding improvement with respect to the conventional VB, nominal KF, VB Sliding Window 
is $91\% $, $13\%, 88\%$ in position ARMSE 
and $45\% $, $12\%$, $36\%$ in velocity, respectively. 
Although the results show performance degradation of both
the VB Sliding Window and the proposed VB MHE when the window length decreases,  
the VB MHE degrades gracefully by providing smaller position and velocity ARMSEs as well as quicker convergence for all values of the window length, especially for low
values of $T$ for which the VB Sliding  Window
may exhibit much worse performance. 

%


\begin{table}[tbp]
	\centering
	\caption{Position ARMSE vs. window length $T$}
	\label{ARMSE-p}
	\begin{tabular}{ccccc}
		\hline
		$T$   &Conventional VB &NKF &VB SW  &VB MHE   \\    
		\hline	 
		 20   & -    &-  &16.2  	&9.4  \\ 
		\hline
		 10   &-	&-	  &19.4    &10.5   \\       
		\hline
		 5    &-	&-	  &65.6	  &15.9 \\       
		\hline
		 4    &-	&-    &172.2   &20.1  \\       
		\hline
		- & 236.7 & 23.2  & - & - \\
		\hline
	\end{tabular}
\vspace{.6cm}
	\centering
	\caption{Velocity ARMSE vs. window length $T$}
	\label{ARMSE-v}
	\begin{tabular}{ccccc}
		\hline
		$T$   &Conventional VB &NKF &VB SW  &VB MHE   \\    
		\hline	 
		 20   &-   &-   &12.0  	&10.7  \\ 
		\hline
		 10   &-	&-	  &12.4    &11.2    \\       
		\hline
		 5    &-	&-	  &15.3   &12.5 \\       
		\hline
		 4    &-	&-    &20.9   &13.4  \\       
		\hline
		 -   &24.6    &15.2   &-  	&-  \\ 	
		\hline
	\end{tabular}
\end{table}

\section{Conclusions}
\label{conclusion}

An adaptive variational Bayes moving horizon estimation method
for state estimation under unknown process and measurement noise covariances 
has been proposed.
Stability analysis  has shown 
that the proposed filter ensures mean-square boundedness of the state estimation error
for any number of VB iterations and any length of the moving window.
Simulation results on a target tracking example have demonstrated the effectiveness 
of the proposed filter.
Future work will focus on consensus adaptive state estimation for networked filtering
with unknown noise covariances as well as the related stability analysis of the distributed filter.

\begin{ack}                               
This work was partly funded by National Natural Science Foundation of China (61627810), 
National Science and Technology Major Program of China (2018YFB1305003), 
and China Scholarship Council. 
\end{ack}


                                 

\bibliographystyle{ifacconf}       
\bibliography{autosam}

\end{document}